# Investigation of Mode Interaction in Optical Microresonators for Kerr Frequency Comb Generation

Yang Liu[1*], Yi Xuan[1,2], Xiaoxiao Xue[1], Pei-Hsun Wang[1], Andrew J. Metcalf[1], Steve Chen[1], Minghao Qi[1,2], and Andrew M. Weiner[1,2]

[1]*School of Electrical and Computer Engineering, Purdue University, 465 Northwestern Avenue, West Lafayette, IN 47907-2035*
[2]*Birck Nanotechnology Center, Purdue University, 1205 West State Street, West Lafayette, Indiana 47907, USA*
*yangliu@purdue.edu*

**Abstract:** Mode interaction in silicon nitride micro-resonators is investigated. We provide clear experimental evidence of mode interaction between two families of transverse modes and demonstrate a link between such interactions and initiation of comb generation in resonators with normal dispersion.

Recently optical comb generation using high quality factor (Q) resonators has been intensively investigated. By tuning a continuous-wave (CW) laser into a cavity resonance, intracavity power builds up and enables additional cavity modes to oscillate through cascaded four-wave mixing (FWM) [1-5]. The comb is generally believed to be generated by modulation instability (MI) of the CW pump mode [6-8]. According to both experimental observation and theoretical analysis, comb generation preferably occurs in resonators with anomalous dispersion. However, comb generation in resonators characterized with normal dispersion has also been observed experimentally [5, 9-11]. Several models have been proposed to describe this phenomenon. Although MI gain is missing in fibers or waveguides with normal dispersion, when it comes to resonators, the detuning provides an extra degree of freedom which enables MI to take place in the normal dispersion regime, hence the generation of the combs [6, 8, 12]. However, following this approach, either hard excitation or precise detuning and pump power relationship needs to be met, making it difficult to realize practically. Alternatively, mode interaction between different family modes has been suggested to be the cause of the comb generation in resonators with normal dispersion [13]. It has yet to be shown experimentally how the two modes can affect each other or whether the combs are generated at the wavelength at which the two modes are interacting. In the anomalous dispersion regime, mode interaction has been reported to affect the bandwidth scaling of the frequency combs [14], but in this case anomalous dispersion is still considered to be the governing factor of the comb generation process. In our report, using normal dispersion resonators, we demonstrate experimentally that the two modes may interact with each other when their resonances come close. In addition, we show the location of one of the primary sidebands from the comb generation will appear at the location of a mode interaction, providing clear experimental evidence that the mode interaction plays a major role in the comb generation for resonators exhibiting normal dispersion.

In this work, the silicon nitride resonator we fabricated has 2 μm × 515 nm waveguide in cross-section. According to the simulation, the silicon nitride waveguide of this dimension will work in normal dispersion regime [11]. The total path length of the resonator is 5.92 mm which corresponds to a free spectra range (FSR) slightly under 25 GHz. Similar to [15], to avoid the stitching error we introduce a finger-shaped structure for the resonator so that it can fit on a single field of our electron beam lithography tool. Figure 1(a) shows a microscope image of the fabricated microresonator. The light is coupled both in and out through lensed fibers which are positioned in U-grooves to improve stability when working at high power. Fiber-to-fiber coupling loss is ~5 dB. The measured transmission spectrum, showing resonances throughout the lightwave C band, is given in Fig. 1(b). If we zoom in the transmission spectrum as shown in Fig. 1(c), resonances of 2 transverse mode families with different depth can be observed. The loaded Q factors at the frequencies shown are ca. $1\times10^6$ and $3\times10^5$ for modes 1 and 2, respectively.

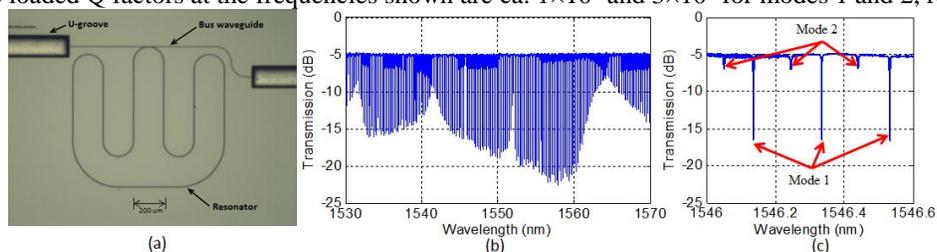

Fig. 1. (a) Microscope image of the silicon nitride resonator. (b) Measured transmission spectrum of the resonator. (c) Zoom-in transmission spectrum showing resonances from different transverse mode families.

We use the frequency-comb-assisted spectroscopy method of [16] to accurately determine the resonance positions and compute the changes in FSR with wavelength to estimate the dispersion for each mode. The measured FSRs are given in Fig. 2(a). The FSRs for the two modes are around 24.8 GHz and 24.4 GHz, respectively. For both modes, the FSR tends to increase with increasing wavelength, which indicates that the resonator is in the normal dispersion regime. However, at several wavelengths when the modes are getting close to each other (1531 nm, 1541 nm and 1561 nm), the FSR of the two modes experience a change and become closer to each other. If we look at the transmission, e.g. in Fig. 2(b), the two modes push each other apart and will not overlap. The large jump of the measured FSR suggests that two modes trade their places: the resonances of the mode with higher FSR which was originally at the left side will switch to the right side. This clearly suggests an interaction between the two modes. A different case for the mode crossing is also observed around 1551 nm given in Fig. 2(c); here there is no obvious change in FSR around the wavelength where the resonances of these modes get close. The bump near 1551.5nm arises when the two resonances overlap, which makes it difficult to determine the exact resonance wavelength for each mode. Looking across the spectrum, we clearly observe that the mode interactions result in a major modification to the local dispersion, even changing the sign of dispersion in some wavelength regions.

For the comb generation experiment, we pumped the micro-resonators with a single CW input at 1.75W for different resonances and recorded the comb spectra. The results are given in Fig. 2(d). When the pumping wavelength is varied, the mode spacing of comb varies from 33 FSR for pumping at 1554 nm to 7 FSR for pumping at 1560 nm. However, one of the two primary sidebands remains at the same or similar position. When we look at the spectral location of the stationary sideband, it coincides with the mode crossing location near 1561 nm. This clearly suggests that mode crossing and mode interaction are major factors in comb generation in this normal dispersion microresonator. We have observed similar results for pumping in other wavelength regions and in other silicon nitride normal dispersion microresonators with FSRs of 37.5 and 75 GHz, which we will report in our presentation.

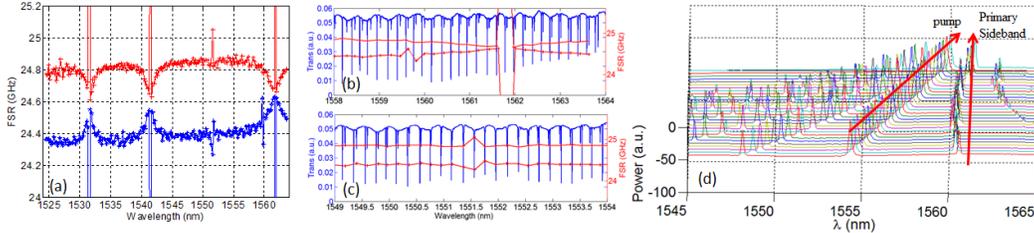

Fig. 2. (a) Measured FSR versus optical wavelength for two transverse modes, plotted in red and blue respectively. (b,c) Transmission (blue) and FSR changes (red) for wavelengths centered near (b) 1561 nm and (c) 1551 nm. (d) Comb spectra for pumping at different resonances.

In summary, we have demonstrated what we believe to conclusive evidence of mode interactions in normal dispersion silicon nitride microresonators and their impact on initiation of comb generation.